\documentclass[onecolumn,groupedaddress,superscriptaddress,floatfix]{revtex4-1}
\usepackage{makeidx}
\usepackage{amssymb}
\usepackage{amsmath}
\usepackage{mathrsfs}
\usepackage{graphicx}
\usepackage{dcolumn}
\usepackage{bm}
\usepackage[center]{subfigure}
\usepackage{color}

\begin{document}

\title{Long-range transverse Ising model built with dipolar condensates in
two-well arrays}
\author{Yongyao Li}
\affiliation{School of Physics and Optoelectronic Engineering, Foshan University, Foshan 528000, China}
\affiliation{Department of Applied Physics, South China Agricultural University,
Guangzhou 510642, China}
\affiliation{Department of Physical Electronics, School of Electrical Engineering,
Faculty of Engineering, Tel Aviv University, Tel Aviv 69978, Israel}
\author{Wei Pang}
\affiliation{Department of Experiment Teaching, Guangdong University of
Technology, Guangzhou 510006, China}
\author{Jun Xu}
\affiliation{Center of Experimental Teaching for Common Basic Courses, South
China Agriculture University, Guangzhou 510642, China} \affiliation{TianQin
Research Center \& School of Physics and Astronomy, Sun Yat-Sen University
(Zhuhai Campus), Zhuhai 519082, China}
\author{Chaohong Lee}
\affiliation{TianQin Research Center \& School of Physics and Astronomy, Sun
Yat-Sen University (Zhuhai Campus), Zhuhai 519082, China}
\author{Boris A. Malomed}
\email{malomed@post.tau.ac.il}
\affiliation{Department of Physical Electronics, School of Electrical Engineering,
Faculty of Engineering, Tel Aviv University, Tel Aviv 69978, Israel}
\author{Luis Santos}
\affiliation{Institut f\"ur Theoretische Physik, Leibniz Universit\"at
Hannover, Appelstr. 2, 30167 Hannover, Germany}

\begin{abstract}
Dipolar Bose-Einstein condensates in an array of double-well potentials
realize an effective transverse Ising model with peculiar inter-layer
interactions, that may result under proper conditions in an anomalous
first-order ferromagnetic-antiferromagnetic phase transition, and nontrivial
phases due to frustration. The considered setup allows as well for the study of
Kibble-Zurek defect formation, whose kink statistics follows that expected
from the universality class of the mean-field one-dimensional transverse Ising model.
Furthermore, random occupation of each layer of the stack leads to random
effective Ising interactions and local transverse fields, that may lead to the
Anderson-like localization of imbalance perturbations.\newline

\end{abstract}

\maketitle

\section{Introduction}

A new generation of experiments with ultra-cold magnetic atoms~\cite%
{Griesmaier2005,Lu2011,Aikawa2012,DePaz2013}, polar molecules~\cite%
{Ni2008,Yan2013,Takehoshi2014,Park2015}, and Rydberg-dressed atoms~\cite%
{Rydberg-Dressed} are starting to reveal novel fascinating physics of
dipolar gases. Whereas in non-dipolar Bose gases inter-particle interactions are
short-range and isotropic, dipolar gases present significant or even
dominant dipole-dipole interactions~(DDI), which are long-range and
anisotropic. As a result, the physics of dipolar gases strongly differs from
that of their non-dipolar counterparts \cite{Lahaye2009,Baranov2012},
featuring effects such as geometry-dependent stability~\cite{Koch2008},
roton-like excitations~\cite{Santos2003,Wilson2008} and roton-dominated
immiscibility \cite{immiscibility,Young2012}, strongly anisotropic vortices
\cite{vortex,YiS2006,Abad2009} and solitons \cite{soliton1,soliton2},
ferrofluidity \cite{ferrofluidity,Saito2009} and anisotropic superfluidity
\cite{Bohn}, striped patterns \cite{stripes}, specific mesoscopic
configurations trapped in triple potential wells \cite{triple}, double- and
triple-periodic ground states in lattices populated by dipolar atoms \cite%
{Tilman}, and the recent discovery of robust quantum droplets~\cite%
{Kadau2016,Ferrier2016}.

Dipolar gases in optical lattices are also remarkably different~from their
non-dipolar counterparts \cite{Lahaye2009,Baranov2012}. Whereas in the absence of
DDI, interparticle interactions in deep lattices reduce to on-site
nonlinearity, the DDI result in inter-site interactions. The latter is true
even for very strong lattices, in which inter-site tunneling vanishes. As a
result, dipolar lattice gases allow for the transport of excitations in the
absence of mass transfer. Recently, spin-like transport was studied in
gases of magnetic atoms~\cite{DePaz2013} and polar molecules~\cite{Yan2013},
where the spin was encoded, respectively, in the electronic spin and in
the rotational degree of freedom. The dipole-induced spin exchange and Ising
interactions result in an effective XXZ Hamiltonian \cite%
{Micheli2006,Gorshov2011}. It has been recently shown that in an imperfectly
filled lattice the dipole-induced spin exchange may result in a peculiar
disorder scenario~\cite{Deng2016}.

\begin{figure}[t]
\includegraphics[width=0.6\columnwidth]{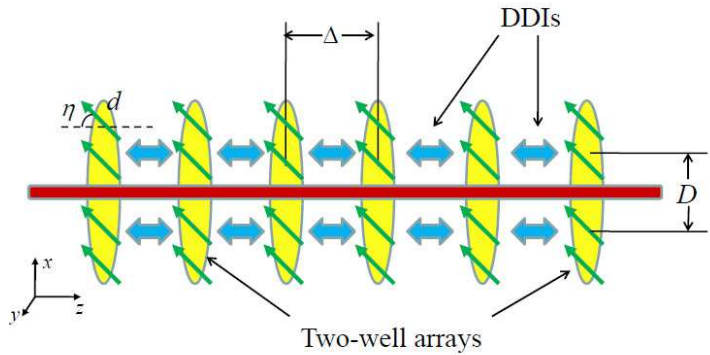}
\caption{(Color online) Sketch of the setup considered: a stack of
two-well arrays of dipolar condensates formed an effective one-dimensional transverse Ising model.}
\label{fig1a}
\end{figure}

In this paper, we discuss a set-up that permits for coding spin-like systems into a spatial degree of freedom of a dipolar Bose-Einstein condensate~(BEC).
The condensate is prepared in a stack of layers of two-well potentials that emulate an effective spin-$1/2$ system~(see Fig.~\ref{fig1a}). This
set-up realizes a transverse Ising model with a peculiar form of long-range interactions that results in an unconventional first-order ferromagnetic-antiferromagnetic transition, as well as
in phases with anomalous periodicities due to magnetic frustration. Since the parameters may be easily changed in real-time the model allows as well for quenching through
second-order phase transitions, as we illustrate for the particular case of a transition from an effective paramagnet into a ferromagnet. We show that
 the associated defect formation follows the Kibble-Zurek~(KZ)~\cite{Kibble1980,Zurek1996,Dziarmaga} scaling expected from the universality class of the mean-field one-dimensional transverse Ising model.
 Furthermore, we show that random layer filling results in an effective disorder
in both the Ising-like interactions and the local transverse field, allowing for the observation of Anderson-like localization of imbalance perturbations.

The paper is organized as follows. In Sec.~\ref{sec:Model} we introduce the set-up and derive the effective long-range transverse Ising model.
Section~\ref{sec:Ground} discusses the corresponding ground-state phases, whereas Sec.~\ref{sec:KZ} comments on the formation of KZ defects.
Section~\ref{sec:Disorder} discusses the effective disorder resulting from random layer filling and the associated
Anderson-localization in the imbalance transport. Finally Sec.~\ref{sec:Conclusions} summarizes our conclusions.


\section{The model}

\label{sec:Model}

We consider in the following a stack of axisymmetric quasi-one-dimensional dipolar BECs (``wires"),
separated along the $z$ direction by a distance $\Delta $, with their axes
oriented along $x$, as shown in Fig. \ref{fig1a}. This configuration may be
readily created by loading the BEC into just one plane of a 2D optical
lattice created in the $yz$ plane. The lattice is assumed deep enough, to
suppress both on-site dynamics along the $y$ and $z$ directions and
tunneling between adjacent condensates. An additional double-well potential $%
U(x)$, with inter-well spacing $D$, is placed along the $x$ axis, while the
atomic dipole moments are parallel to the $xz$ plane, forming angle $\eta $
with the $z$ axis. The system is described by a set of coupled one-dimensional
Gross-Pitaevskii~(GP) equations:
\begin{equation}
i\hbar\frac{\partial}{\partial t}\psi_{n}(x,t)=\left[ -\frac{\hbar^2}{2m}\frac{%
\partial ^{2}}{\partial x^{2}}+U(x)+\mathcal{V}_{n}(x)\right] \psi _{n}(x,t),
\label{eq:GP}
\end{equation}%
with $m$ the particle mass, $\psi _{n}(x,t)$ the axial wave function at site $n$, and
\begin{equation}
\mathcal{V}_{n}(x)\equiv \int_{-\infty }^{+\infty }dx^{\prime }\left[\sum_{n'}V_{n'-n}(x-x')+g_{1D}\delta (x-x')\delta _{n'n}\right] |\psi _{n'}(x^{\prime},t)|^{2}.
\end{equation}
The contact interactions are characterized by $g_{1D}=\frac{2\hbar ^{2}a}{ml^{2}}$, with $a$ the scattering length, and $l$ the
effective oscillator length associated to the on-site confinement in the $yz$
plane. $V_{n'-n}(x)$ is the DDI between dipoles
placed $n'-n$ sites apart and separated by an axial distance $x$. The kernel $V_{n'-n}(x)$
is the Fourier transform of
\begin{equation}
\tilde{V}_{n'-n}(k_{x})=\int \frac{dk_{y}}{2\pi }\int \frac{dk_{z}}{2\pi }%
\tilde{V}_{\mathrm{dd}}(\vec{k})e^{-ik_{z}(n'-n)\Delta
}e^{-(k_{y}^{2}+k_{z}^{2})l^{2}/2},
\end{equation}%
with $\tilde{V}_{\mathrm{dd}}(\vec{k})=\frac{4\pi }{3}d^{2}\left[ 3|\vec{k}%
|^{-2}(k_{z}\cos \eta +k_{x}\sin \eta )^{2}-1\right] $ and $d$ the dipole
moment.

For a sufficiently tight $U(x)$ potential, we may employ a simplified two-mode scenario in which
only the two lowest eigenstates of $U(x)$ participate in the dynamics, $(R(x)\pm L(x))/\sqrt{2}$, where $R(x)$~($L(x)$) denote
the wave functions at the right~(left) well. We may then express
$\psi _{n}(x,t)=a_{n}(t)L(x)+b_{n}(t)R(x)$. The two wells are coherently coupled by a hopping rate $J$~\cite{footnote-quantum}.
Under these conditions, the coupled GP equations~(\ref{eq:GP}) reduce to
\begin{eqnarray}
i\dot{a}_{n}(t) &=&-Jb_{n}(t)+\tilde{\mu}_{n}(t)a_{n}(t),  \label{anbn1} \\
i\dot{b}_{n}(t) &=&-Ja_{n}(t)+\tilde{\mu}_{n}^{\prime }(t)b_{n}(t),
\label{anbn2}
\end{eqnarray}%
where
\begin{eqnarray}
\tilde{\mu}_{n}(t) &\equiv &\sum_{n'}\left[ \left( \tilde{F}_{0}\left(
n'-n\right) +U_{0}\delta _{n'n}\right) \left\vert a_{n'}(t)\right\vert ^{2}\right.
\notag \\
&&\left. +\tilde{F}_{0}\left( n'-n\right) \left\vert b_{n'}(t)\right\vert ^{2}%
\right] N_{n'},
\end{eqnarray}%
$\tilde{\mu}^{\prime }$ is defined with $a_{n}\rightleftarrows b_{n}$, $%
U_{0}\equiv g_{1D}\int_{-\infty }^{+\infty }dx|R(x)|^{4}$, $\tilde{F}%
_{0}(n'-n)\equiv \int_{-\infty }^{+\infty }dx\int_{-\infty }^{+\infty
}dx^{\prime }V_{n'-n}(x-x^{\prime })|R(x)|^{2}|R(x^{\prime })|^{2}$ denotes
the interaction between right wells at two sites placed $n\Delta $
apart~(or equivalently between left wells), $\tilde{F}_{1}(n'-n)\equiv
\int_{-\infty }^{+\infty }dx\int_{-\infty }^{+\infty }dx^{\prime
}V_{n'-n}(x-x^{\prime })|A(x)|^{2}|B(x^{\prime })|^{2}$ is
the interaction between left and right wells, and $N_{n}$ denotes the number
of particles in the $n$-th wire. Since we assume a vanishing inter-site hopping, $%
N_{n}$ is conserved, and $|a_{n}|^{2}+|b_{n}|^{2}=1$. In the following, we assume that the
scattering length is tuned by means of Feshbach resonances, so that $U_{0}=%
\tilde{F}_{1}(0)-\tilde{F}_{0}(0)$. In this way, the on-site (dipolar plus
contact) interactions cancel, allowing us to concentrate on
the non-trivial dynamics arising from the inter-layer DDI. Finally, although the exact form of $%
\tilde{F}_{0}(n'-n)$ and $\tilde{F}_{1}(n'-n)$ may be evaluated exactly, we may
further simplify the model by considering a point-like approximation that
yields
\begin{gather}
\frac{F_{0}(n'-n)}{d^2/\Delta^3}=\frac{1-3\cos ^{2}\eta }{(n'-n)^{3}},  \notag \\
\frac{F_{1}(n'-n)}{d^2/\Delta^3}=\frac{1}{\left[ (n'-n)^{2}+\left( D/\Delta \right) ^{2}\right]
^{3/2}}  \notag \\
-\frac{3\left[ (n'-n)\cos \eta +\left( D/\Delta \right) \sin \eta \right] ^{2}}{%
\left[ (n'-n)^{2}+\left( D/\Delta \right) ^{2}\right] ^{5/2}}.  \label{F0F1}
\end{gather}%
The exact evaluation of $F_0$ and $F_1$ may modify these values, especially for nearest-neighboring wires for which the finite wave packet spreading
may be significant compared to the inter-site spacing, but our results would remain qualitatively unaffected.


\section{Ground-state phases}

\label{sec:Ground}
Interestingly, the system under consideration is equivalent to a spin-$1/2$ transverse Ising model with peculiar long-range Ising interactions given by the Hamiltonian
\begin{eqnarray}
H=-J\sum_{n}N_n S^{x}_{n}+{1\over2}\sum_{n,n'}N_nN_{n'} V_S(n-n')S^{z}_{n}S^{z}_{n'},
\label{Ham2}
\end{eqnarray}
where $-J$ plays the role of an effective transversal magnetic field, $V_S(n'-n)\equiv\left[ F_{0}(n'-n)-F_{1}(n'-n) \right]/2$ characterizes an effective Ising-like coupling, and we have introduced
the effective spin components $S^{x}_{n}=a^{\ast}_{n}b_{n}+\mathrm{c.c}$ and  $S^{z}_{n}=|b_{n}|^{2}-|a_{n}|^{2}$.

\begin{figure}[t]
\subfigure[]{\includegraphics[width=0.33\columnwidth]{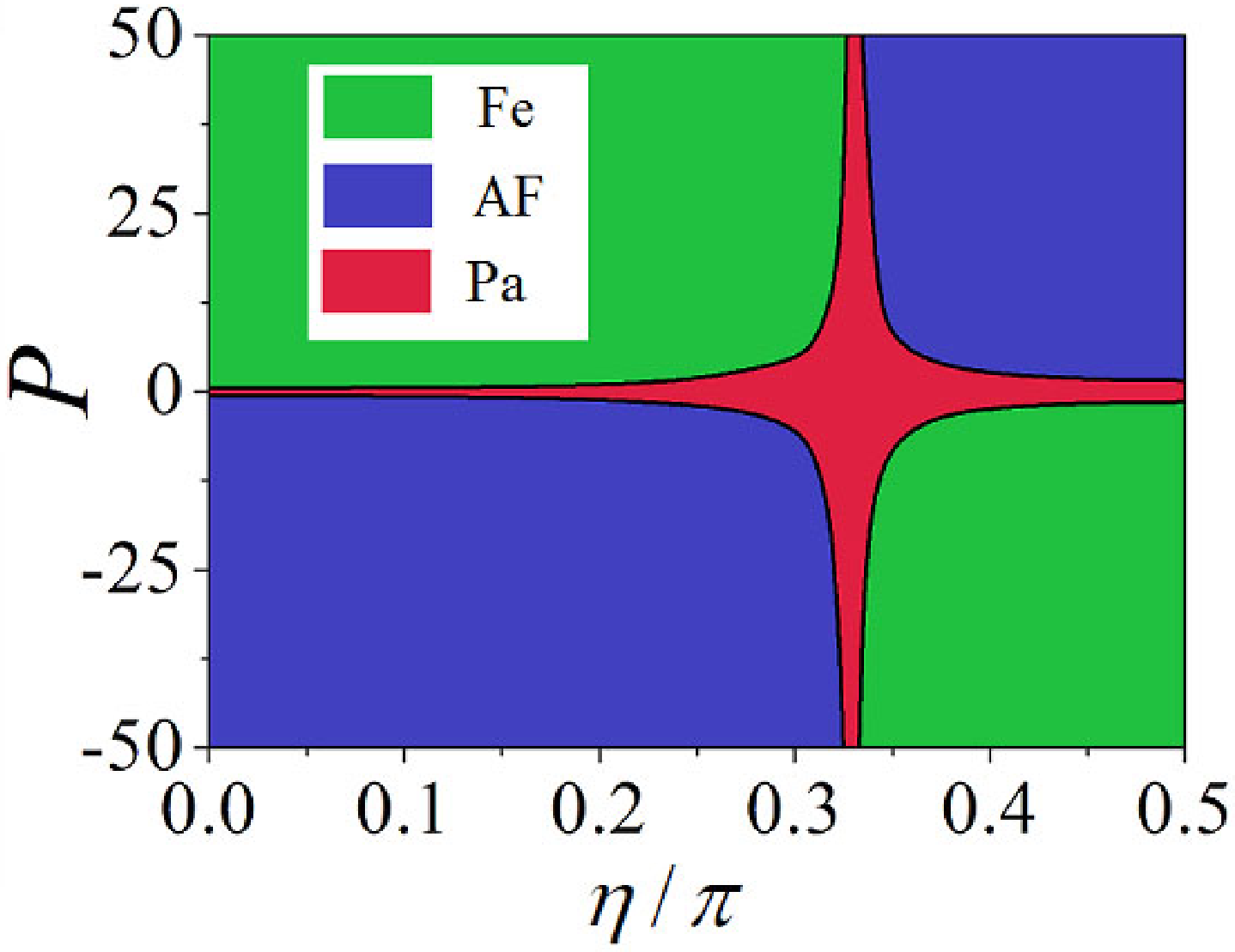}} %
\subfigure[]{\includegraphics[width=0.33\columnwidth]{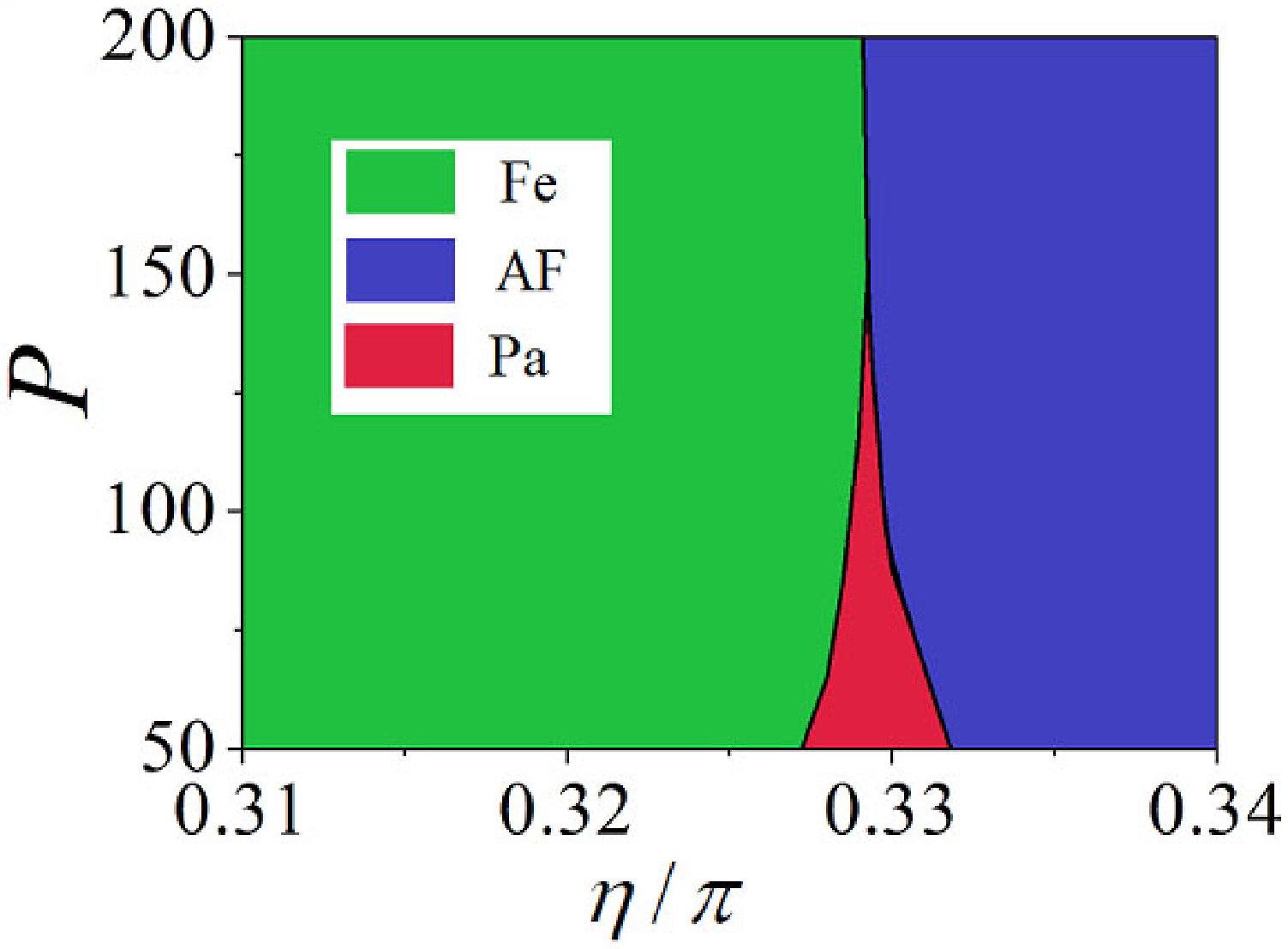}} %
\subfigure[]{\includegraphics[width=0.33\columnwidth]{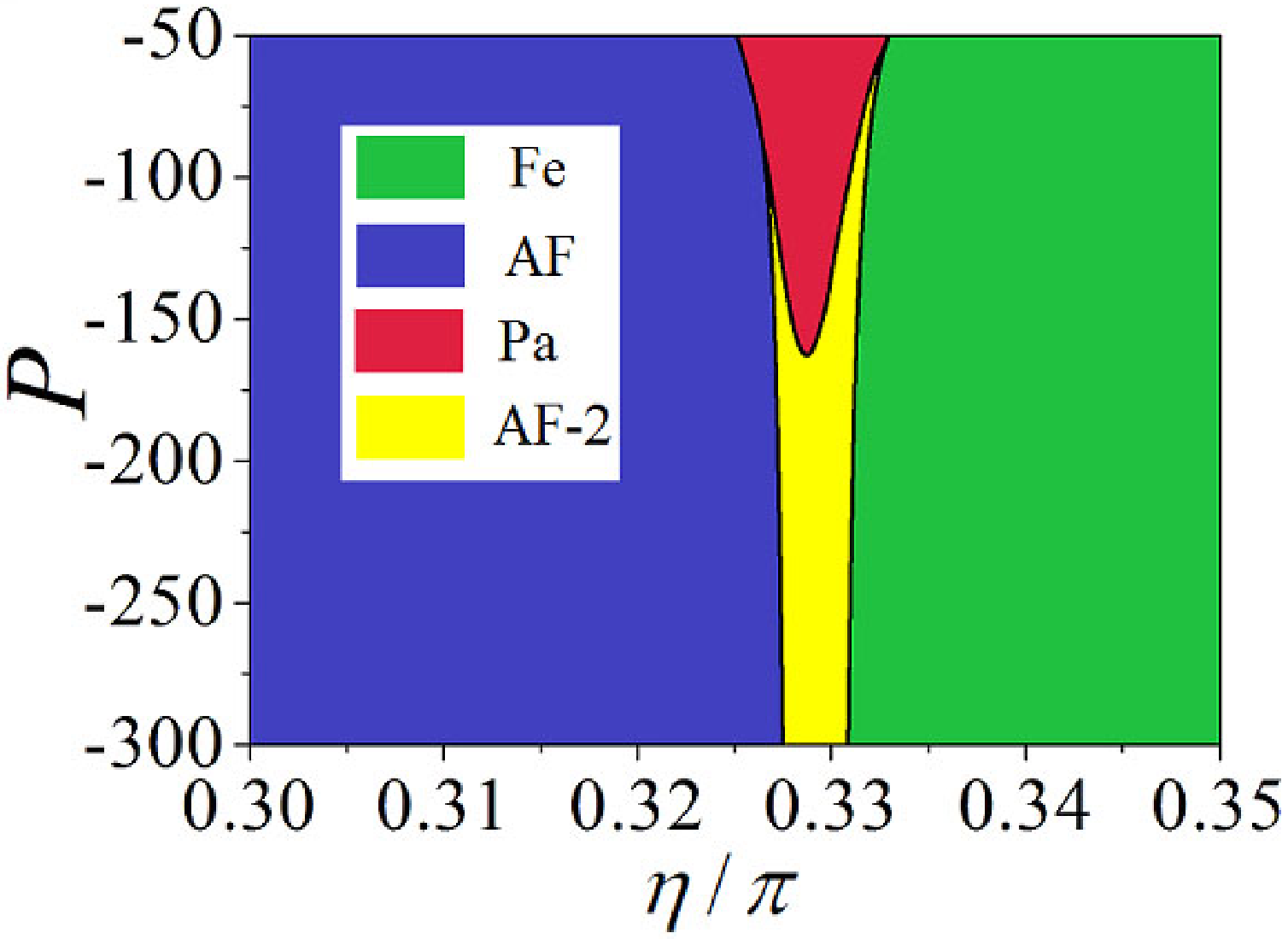}} %
\subfigure[]{\includegraphics[width=0.33\columnwidth]{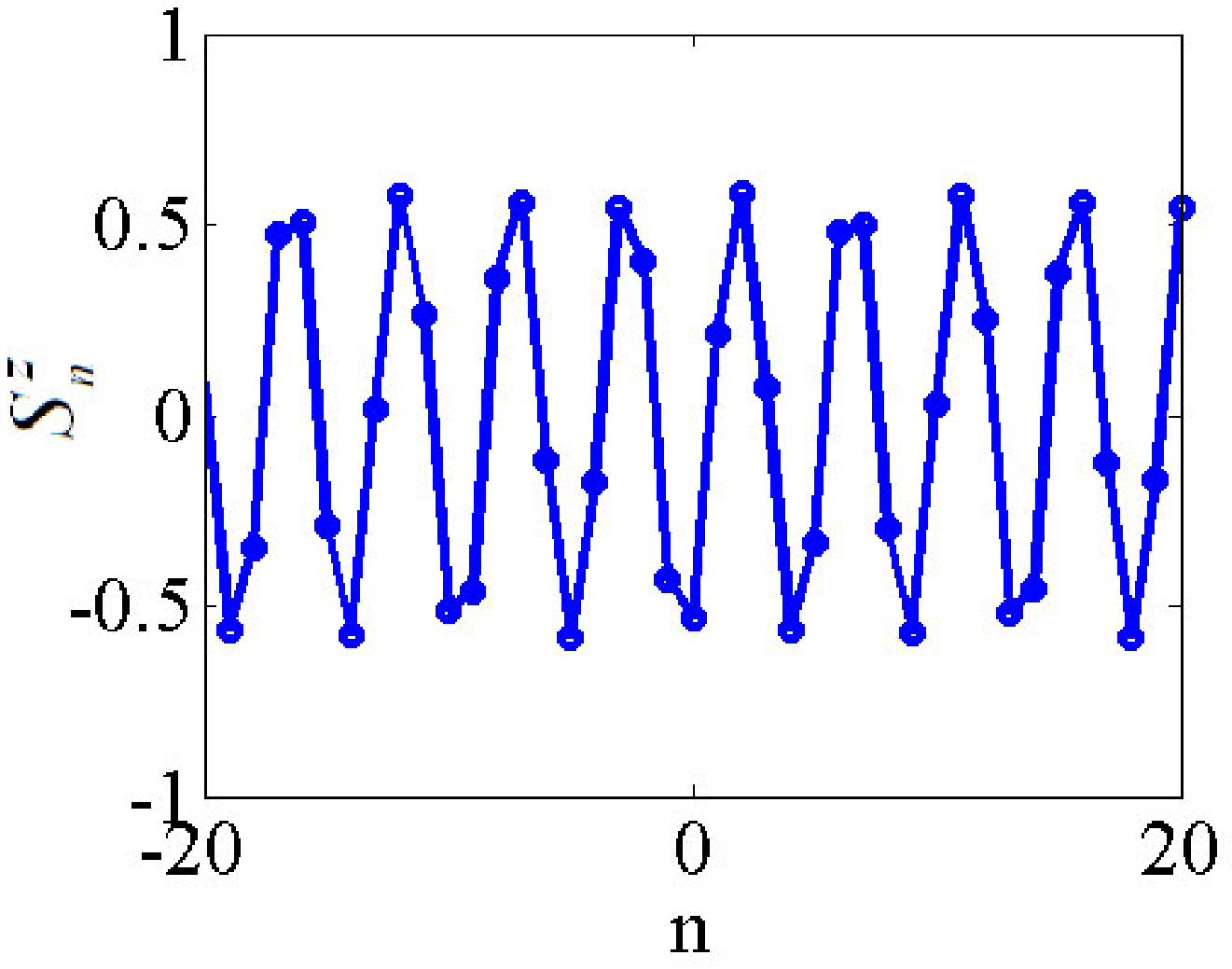}}
\caption{(Color online) (a) Ground-state phase diagram as a function of
the DDI strength $P=Nd^2/J\Delta^3$ and the angle
$\protect\eta $. (b) Vicinity of $\protect\eta _{\mathrm{cr}}$ for
$P>0$, showing the first-order F-AF transition. (c) Vicinity of $\protect%
\eta _{\mathrm{cr}}$ for $P<0$, showing the appearance of the AF-2 phase.
(d) AF-2 phase for $\protect\eta =\protect\eta _{\mathrm{cr}%
}$ and $P=-180$; note the formation of an AF order with a periodicity of approximately five wires.}
\label{phasesdiagram}
\end{figure}

At this point, we assume that all layers are equally populated, $N_{n}\equiv
N$~(we relax this condition in Sec.~\ref{sec:Disorder}). We fix the
hopping rate as the energy unit, i.e. $J=1$, and also
set $D=1$. The strength of the DDI is characterized by the parameter $P=Nd^2/J\Delta^3$, which plays a key role in the discussion below.
For the particular case of dysprosium atoms with an inter-wire separation of $\Delta=1\mu$m, $d^2/\Delta^3\sim 1$ Hz, and hence for $N=10^3$--$10^4$ atoms,
$Nd^2/\Delta^3=1$--$10$ kHz. The corresponding value of $P$ depends on $J$, which is controlled by the barrier of the two-well potential $U(x)$.
For typical values $J\sim 100$ Hz, $P\sim 100$ may be hence readily reached.

The ground-state phase diagram of the system~\cite{footnote-quantum}, presented in Fig.~\ref%
{phasesdiagram}, is obtained numerically from the imaginary-time evolution
of Eqs.~\eqref{anbn1} and~\eqref{anbn2}. If $\eta $ is such that $V_{S}(n'-n)<0$, the Ising interaction is
ferromagnetic. For $P/J=0$ the ground-state of the system is given by a spin oriented along the transversal magnetic field, i.e., along $x$-axis, and hence a solution with zero imbalance $S^{z}_{n}=0$ is favored.
This ground state corresponds to the paramagnetic~(Pa) phase. For a sufficiently large $P>P_{\mathrm{cr}}$~($P_{\mathrm{cr}}\simeq0.45J$ for $\eta=0$),
the system experiences a second-order phase transition into a ferromagnetic~(F) phase,
characterized by a full imbalance, either to the $R$ or to the $L$ well.
 At $\eta_{\mathrm{cr}}\approx0.33\pi$, $V_S(1)=0$ and hence the nearest-neighbor~(NN) interaction changes the sign.
As a result for $\eta>\eta_{\mathrm{cr}}$ at a sufficiently large $P/J$ the system enters an Ising anti-ferromagnetic~(AF) phase,
characterized by a staggered imbalance between neighboring wires.
The situation is obviously reversed for $P<0$~(which may be achieved by means of a rotating magnetic field~\cite{Giovanazzi2002}),
and the Pa-AF transition occurs for  $\eta<\eta_{\mathrm{cr}}$ and Pa-F for $\eta>\eta_{\mathrm{cr}}$.

The situation is particularly noteworthy in the vicinity of $\eta _{\mathrm{cr}%
}$. Whereas for $P<P_{\mathrm{cr}}(\eta _{\mathrm{cr}})\approx 147$, the F
and AF phases remain separated by a Pa phase, for $P>P_{\mathrm{cr}}(\eta _{%
\mathrm{cr}})$ there is a first-order F-AF phase transition, see Fig. \ref%
{phasesdiagram}(b). The reason for this change is that, when $%
|V_{S}(1)|<|V_{S}(2)|$ at $\eta =\eta _{\mathrm{cr}}$, $V_{S}(2)$ remains
negative, i.e., $V_{S}(2)$ favors ferromagnetism between
next-nearest-neighbors~(NNN). This is both compatible with N\'{e}el ordering
and with a fully ferromagnetic state. The only difference between these two
choices is the orientation between NN, which steeply
changes when $V_{S}(1)$ changes its sign. This is remarkably different from
the usual situation in NN Ising models, with $V_{S}(n>1)=0$, in which
the change of the sign of $V_{S}(1)$ implies vanishing interactions, and
hence the Pa phase always separates the F and AF phases. It is also different
from the standard version of the long-range transverse Ising model induced
by dipolar interactions, i.e., $V_{S}(n'-n)=V_{0}/(n'-n)^{3}$. In that case, the
change of the NN coupling at $V_{0}=0$ from F to AF also implies vanishing
of all interactions, and hence the existence of an intermediate Pa phase.
Here, when $P>P_{\mathrm{cr}}(\eta _{\mathrm{cr}})$, $V_{S}(1)$ is
negligible, and $V_{S}(2)$ dominates. Such a dominating ferromagnetic NNN\
coupling allows for a direct first-order transition between F and AF as a function of $\eta$.

A similar competition at $P<0$ results in magnetic frustration. In the vicinity of $%
\eta _{\mathrm{cr}}$, when $|V_{S}(1)|<|V_{S}(2)|$, one has $V_{S}(2)>0$.
Under these conditions, the system experiences frustration, as AF NNN
interactions are now incompatible with the small F or AF NN coupling. As a result, in the vicinity of $\eta _{\mathrm{cr}}$,
a new phase~(AF-2) develops, see Fig.~\ref{phasesdiagram}(c), with an approximate five-site-periodic
modulation of the imbalance, see Fig. \ref{phasesdiagram}(d).


\section{Kibble-Zurek scenario}

\label{sec:KZ}

As shown in Sec.~\ref{sec:Ground} varying $P$ and/or $\eta$ permits
accessing various second-order phase transitions. We note that both parameters
may be modified in real time. In particular $P$ may be readily modified by
altering the barrier between the two wells, since the latter controls the value of $J$.
This provides the possibility of quenching in real time through the second-order phase transitions
of Fig.~\ref{phasesdiagram}. Quenching at a finite speed is expected to induce
defects due to the Kibble-Zurek~(KZ) mechanism~\cite{Kibble1980,Zurek1996,Dziarmaga} .

\begin{figure}[t]
\subfigure[]{\includegraphics[width=0.3\columnwidth]{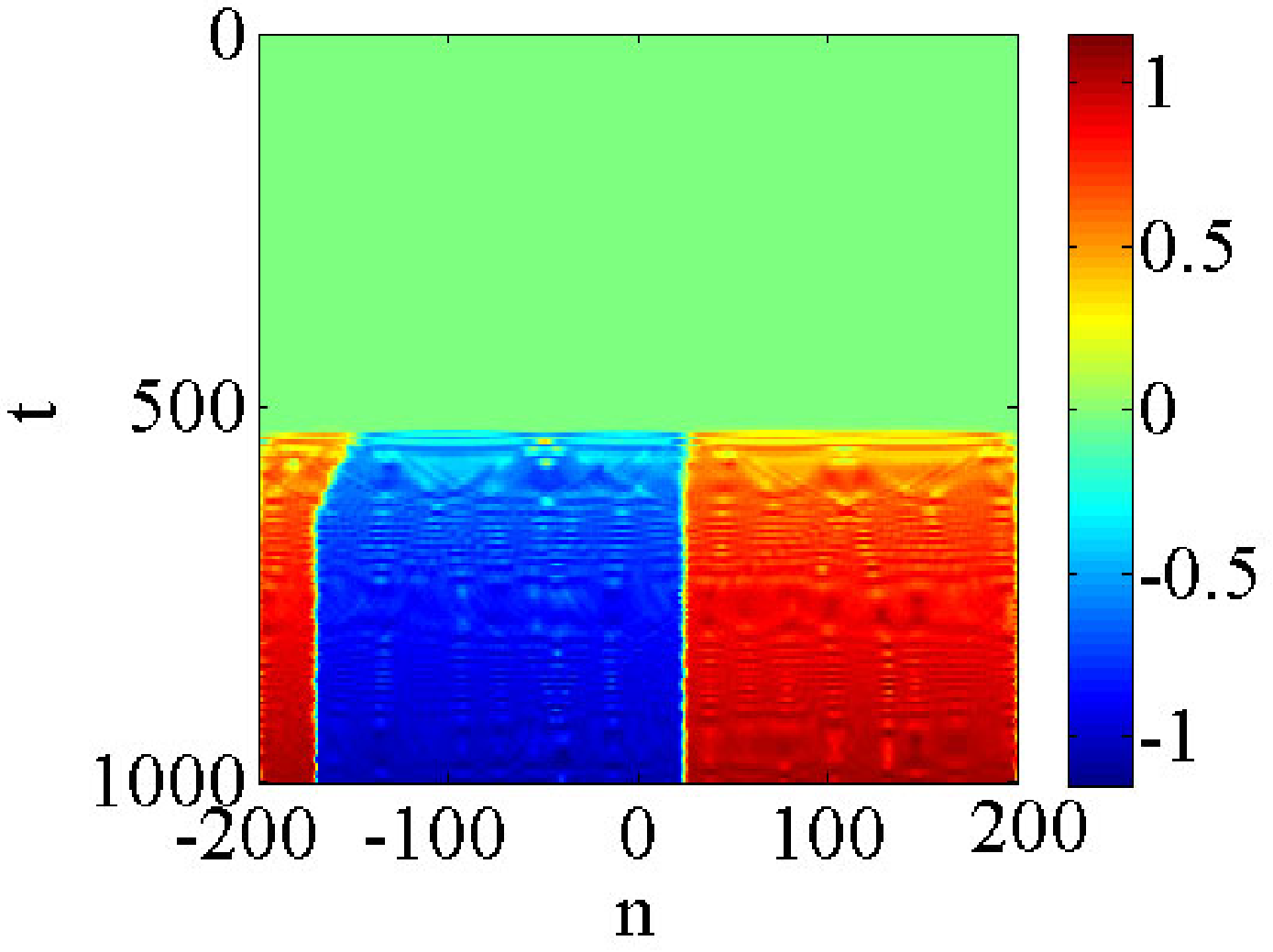}} %
\subfigure[]{\includegraphics[width=0.3\columnwidth]{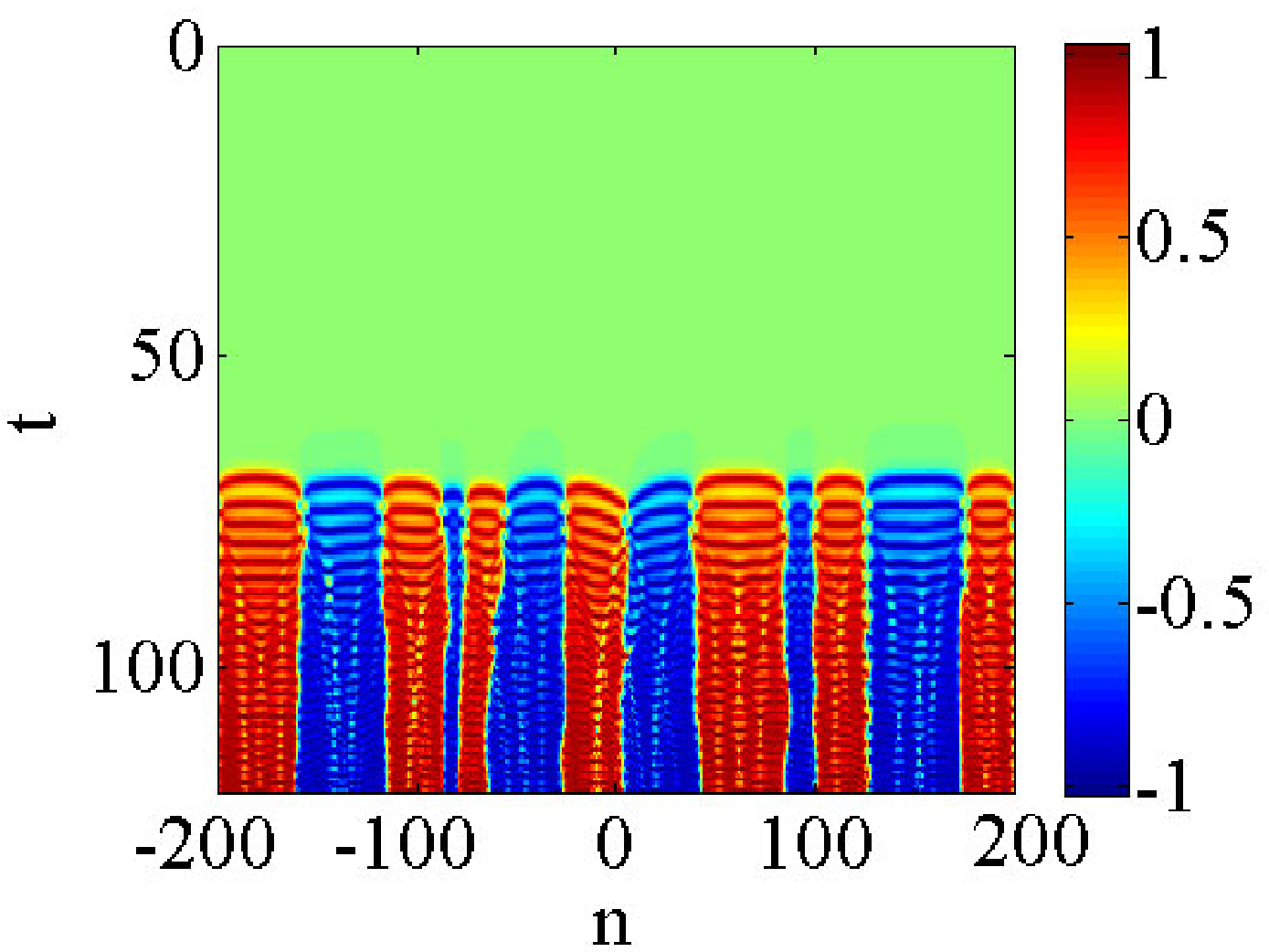}} %
\subfigure[]{\includegraphics[width=0.3\columnwidth]{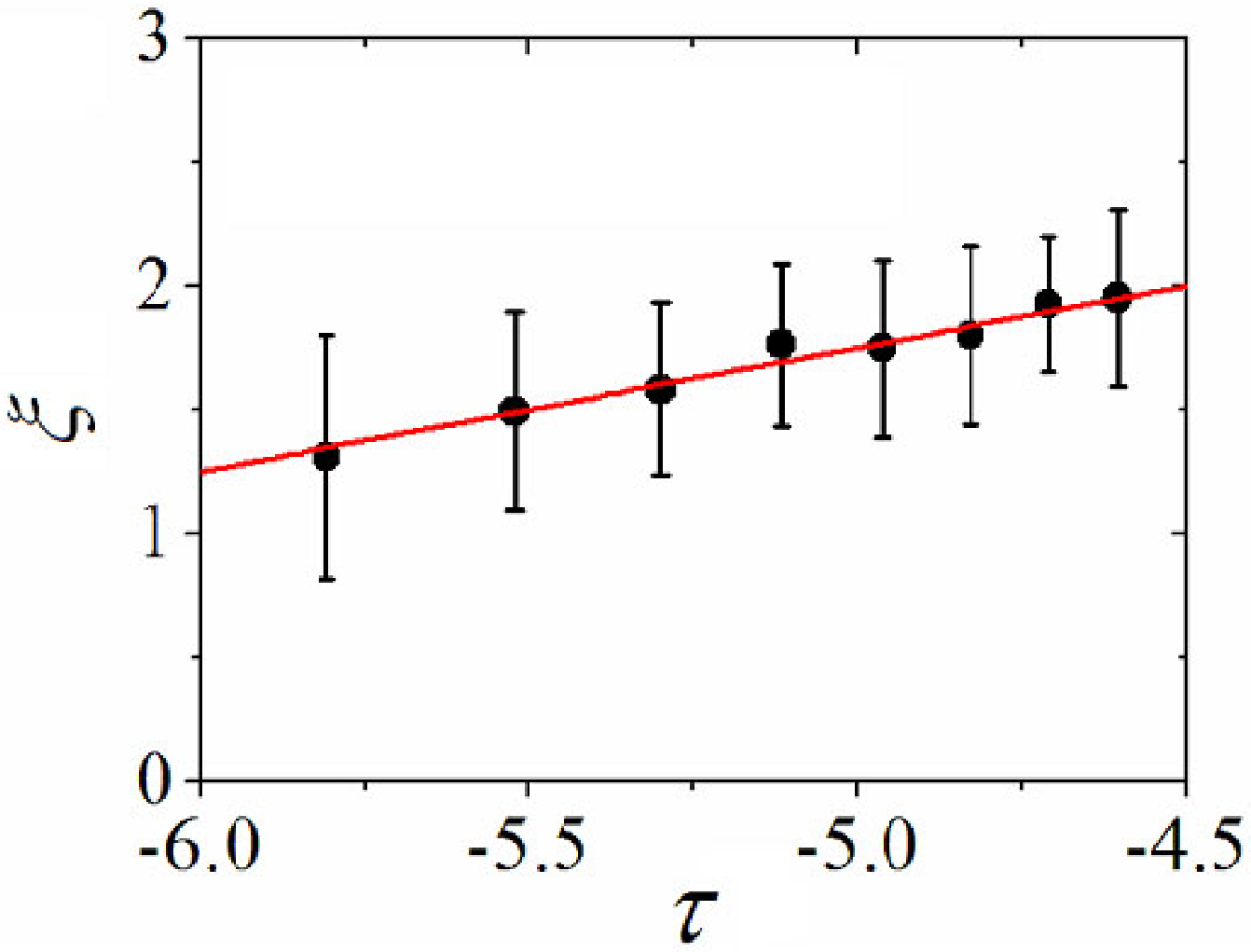}}
\caption{(Color online) Pattern formation in the imbalance
distribution as a function of time for $\eta=0$ and a linear ramp $P(t)=\gamma t$ with a ramp speed $\protect\gamma =10^{-3}$~(a) and $%
10^{-2}$~(b); (c) $\protect\xi =\ln N_{\mathrm{D}}$, where $N_{\mathrm{D}}$
is the number of domains, as a function of $\protect\tau \equiv \ln \protect%
\gamma $, the results being best fitted by $\protect\xi =4.25+0.5\protect%
\tau $, which implies that $N_{\mathrm{D}}\propto \protect\gamma ^{1/2}$, as expected from the KZ scaling.}
\label{KZexp}
\end{figure}

We illustrate this possibility with the particular case of the Pa-F transition. Increasing $P$ for $\eta=0$ eventually
quenches from the fully balanced Pa phase into the F one. As a result, the
system develops F domains, i.e. regions with total imbalance biased to
the $R$ or $L$ sites, separated by a domain wall~(kink). In our
simulations of Eqs.~\eqref{anbn1} and~\eqref{anbn2}, we consider a balanced
input with a slight random imbalance and relative phase perturbation: $%
a_{n}=(0.5-\varepsilon \,\mathrm{\rho}_{1})^{1/2}\,\exp (i\varepsilon \,\mathrm{%
\rho}_{2})$ and $b_{n}=(0.5+\varepsilon \,\mathrm{\rho}_{1})^{1/2}\,\exp
(-i\varepsilon \,\mathrm{\rho}_{2})$, where $-1<\mathrm{\rho}_{1,2}<1$ are two
sets of random numbers, and $\varepsilon \ll 1$~($10^{-6}$ in our
calculations) is the strength of the randomness. This small randomness
mimics slight imperfections that seed the domain-wall formation. We then
impose a linear ramp, $P(t)=\gamma t$, with different ramp speeds $\gamma $.
Typical numerical results for two values of $\gamma $ are displayed in Figs.~%
\ref{KZexp}(a) and~(b). As expected, the number of kinks increases with the ramp
speed $\gamma$ when crossing the transition. From a large number of random
realizations~(up to $50$ different sets of $\rho_{1,2}$), we extract, for each
value of $\gamma $, statistics of the number of the domain walls, $N_{D}$.
Figure~\ref{KZexp}~(c) depicts $\ln (N_{D})$ as a function of $\ln (\gamma )$%
, showing that $N_{D}\sim \gamma ^{1/2}$. The later follows the known KZ
scaling, $N_{\mathrm{D}}\sim \gamma ^{\nu /(\nu z+1)}$, where $\nu =1$ and $%
z=1$ are the critical static and dynamical exponents for the mean-field
one-dimensional transverse Ising model \cite{Dziarmaga}.

\section{Imbalance transport in the presence of random fillings}

\label{sec:Disorder}

The coupling between layers in Eq.~\eqref{Ham2} crucially depends on the
number of particles in each layer. This opens interesting possibilities for
the study of excitation transport --- in particular, localization due to
random interactions, rather than due to random hopping~(we recall that mass
transport between wires is suppressed). We consider a
randomized distribution of the number of particles in each wire, $%
N_{n}/N=1+\varepsilon R_{n}$, where $-1<R_{n}<1$ are random numbers, and $%
\varepsilon \in \lbrack 0,1]$ determines the strength of the randomness.
Such random distributions may be created by abruptly growing the lattice on top of a trapped BEC.
Note that the random population in
each wire translates into a random inter-wire interaction in Eq.~\eqref{Ham2}%
, which may significantly affect the transport of imbalanced excitations.

\begin{figure}[t]
\subfigure[]{\includegraphics[width=0.33\columnwidth]{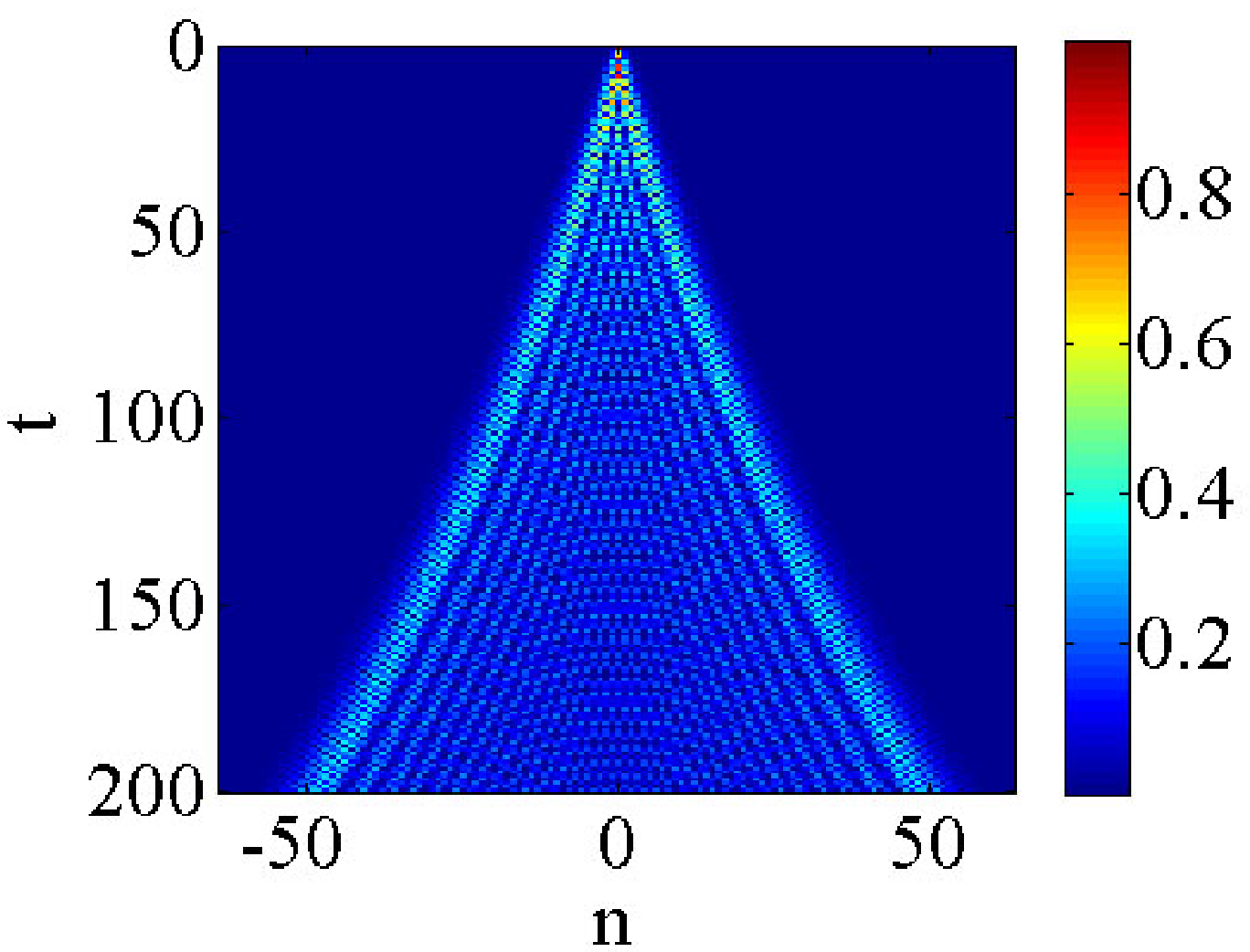}} %
\subfigure[]{\includegraphics[width=0.33\columnwidth]{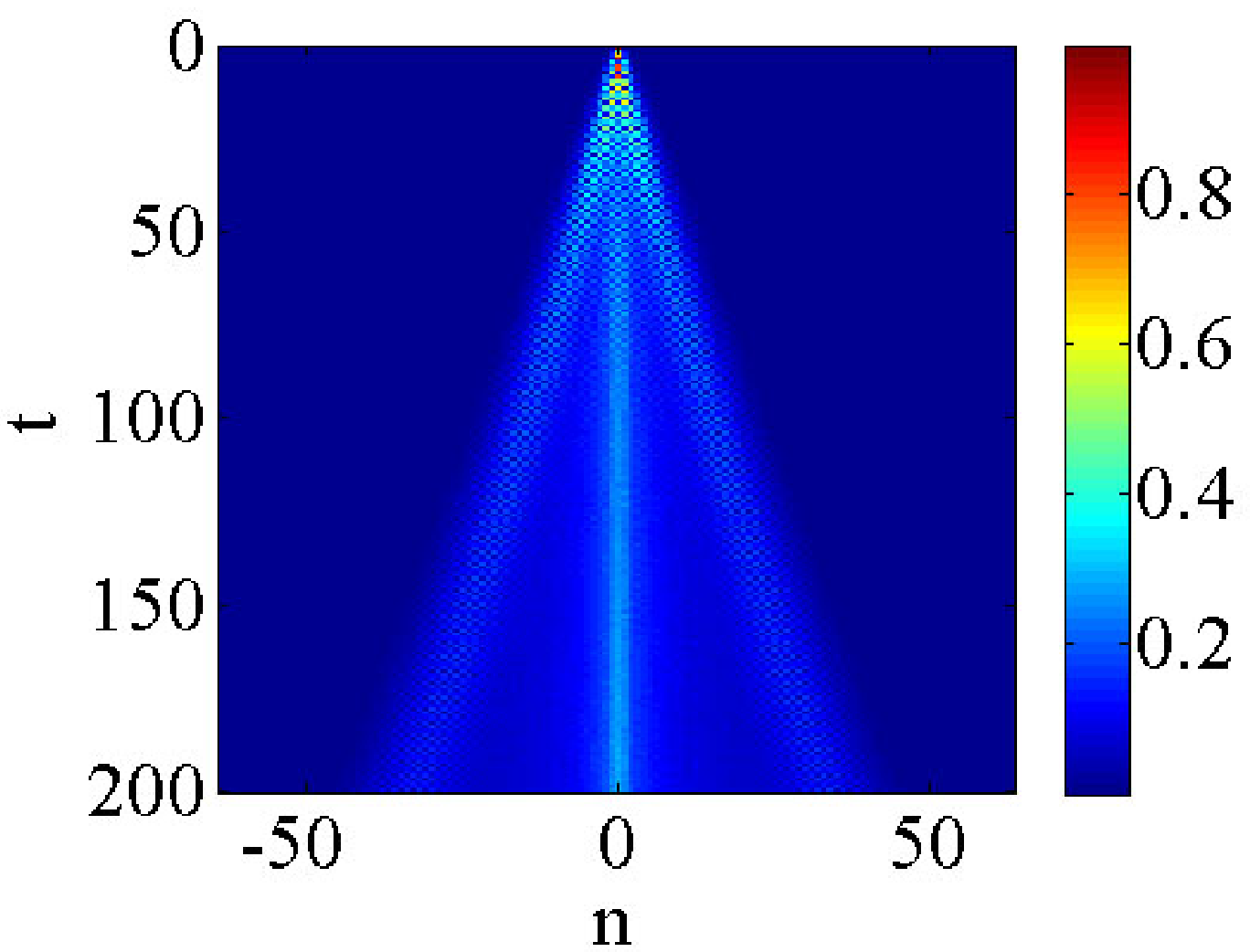}} %
\subfigure[]{\includegraphics[width=0.33\columnwidth]{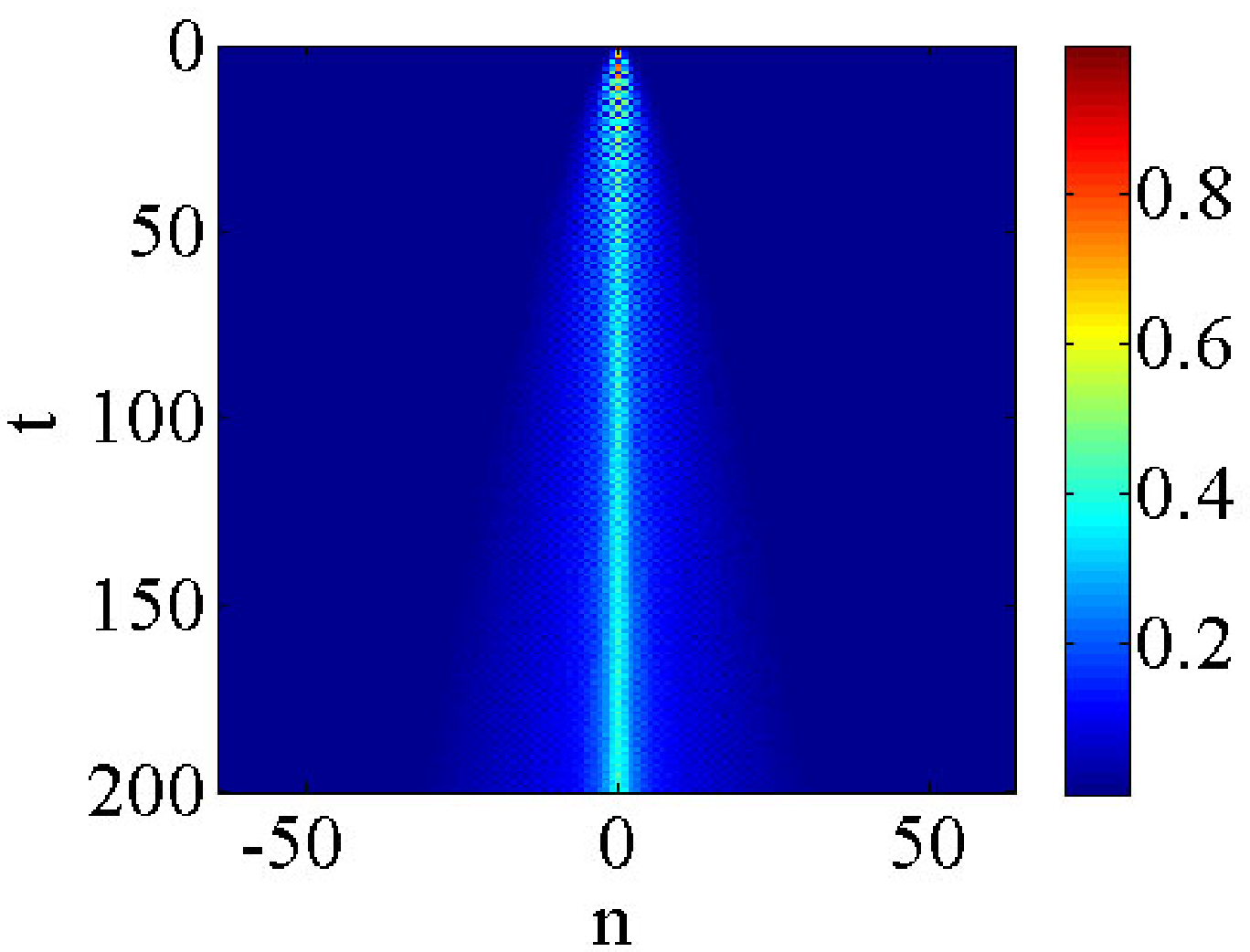}} %
\subfigure[]{\includegraphics[width=0.33\columnwidth]{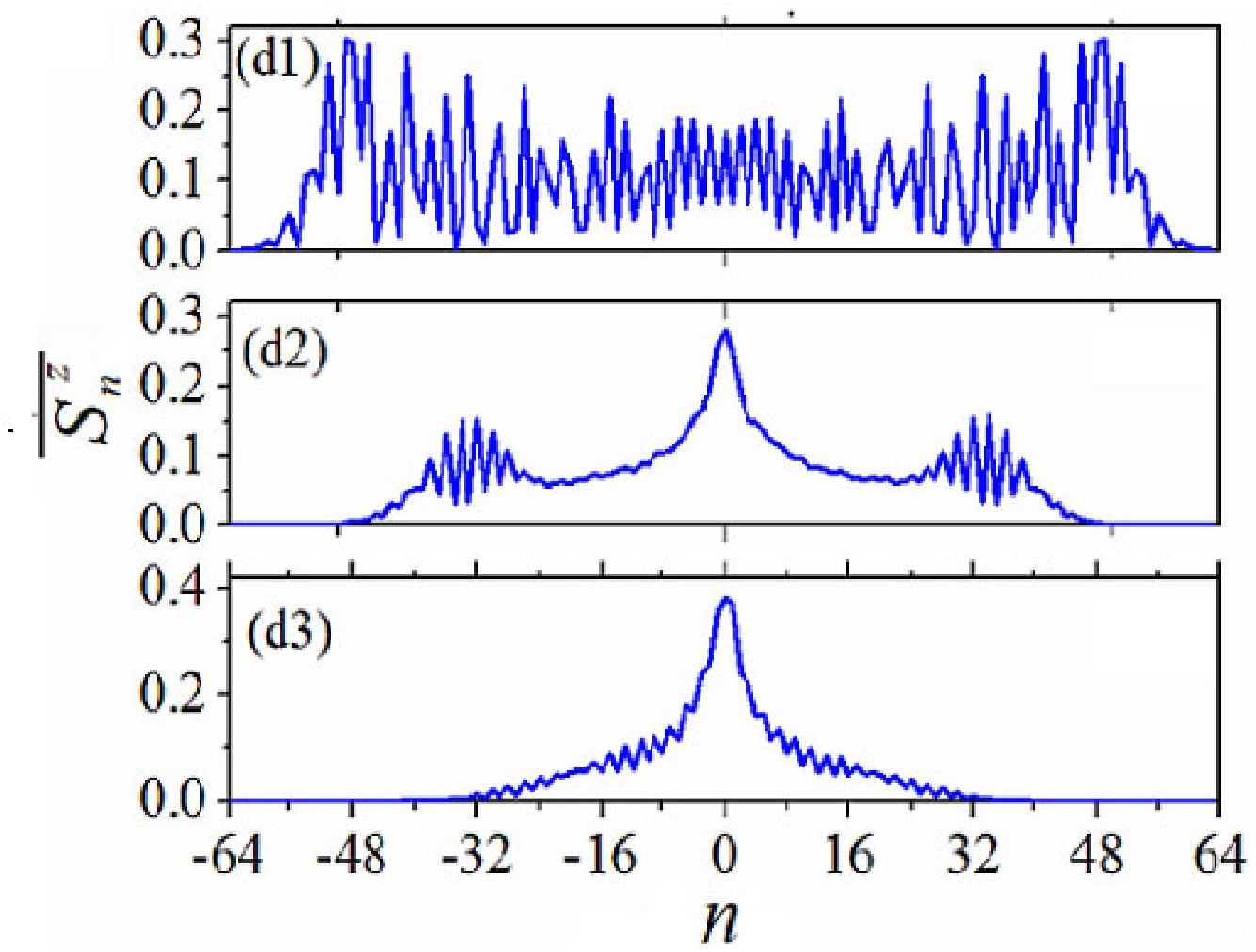}}
\caption{(Color online) Averaged imbalance distribution $\overline{S_{j}^{z}}%
(t)$ for an initial imbalance perturbation localized at the site $n=0$, with $\eta=0$, $P=0.1$,
and  a disorder strength~(see text) $\protect\varepsilon =0$~(a), $0.4$~(b), and $0.8$~(c).
Figure (d) shows from top to bottom $\overline{S_{n}^{z}}(t=200)$ for $\protect%
\varepsilon =0$, $0.4$, and $0.8$. }
\label{totalwave}
\end{figure}

\begin{figure}[t]
\subfigure[]{\includegraphics[width=0.33\columnwidth]{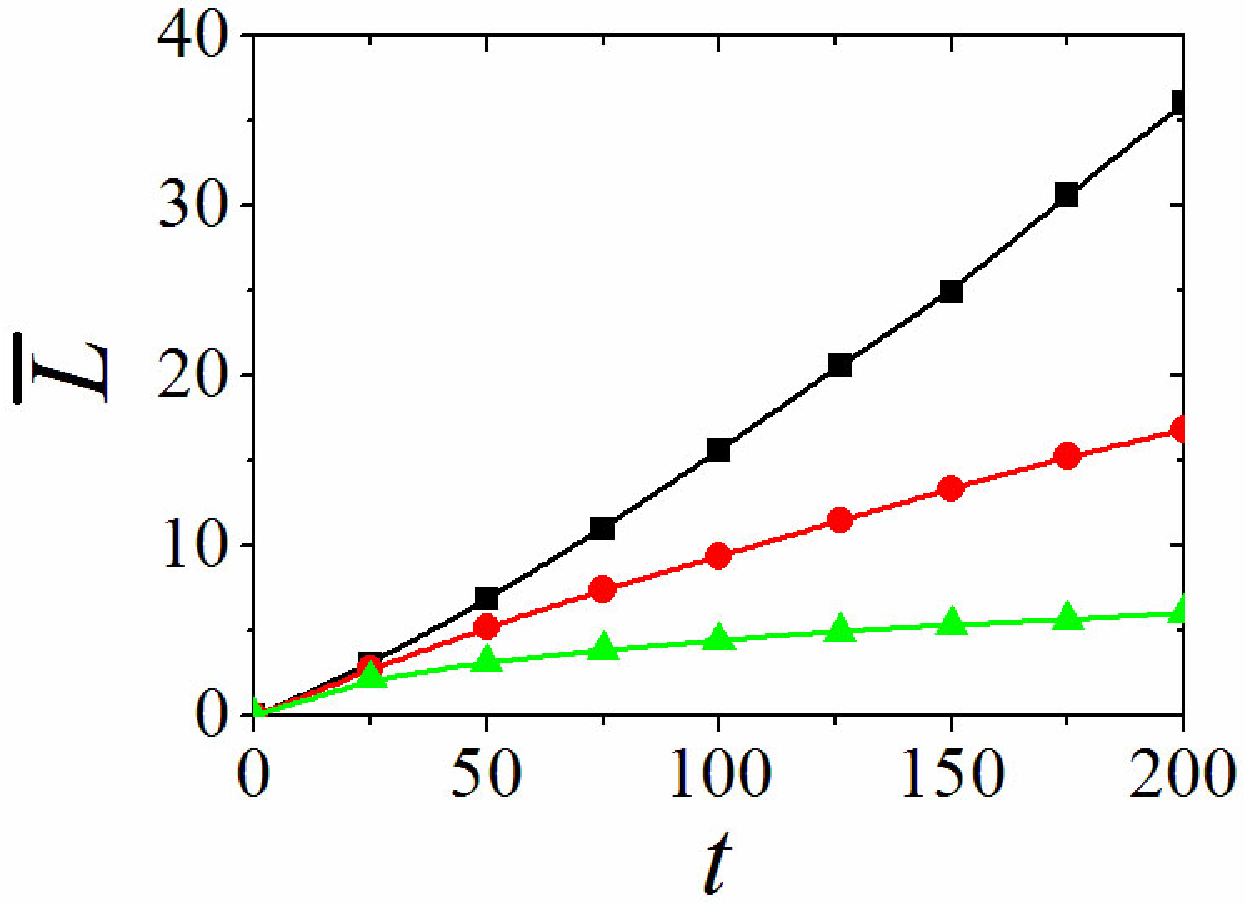}} %
\subfigure[]{\includegraphics[width=0.33\columnwidth]{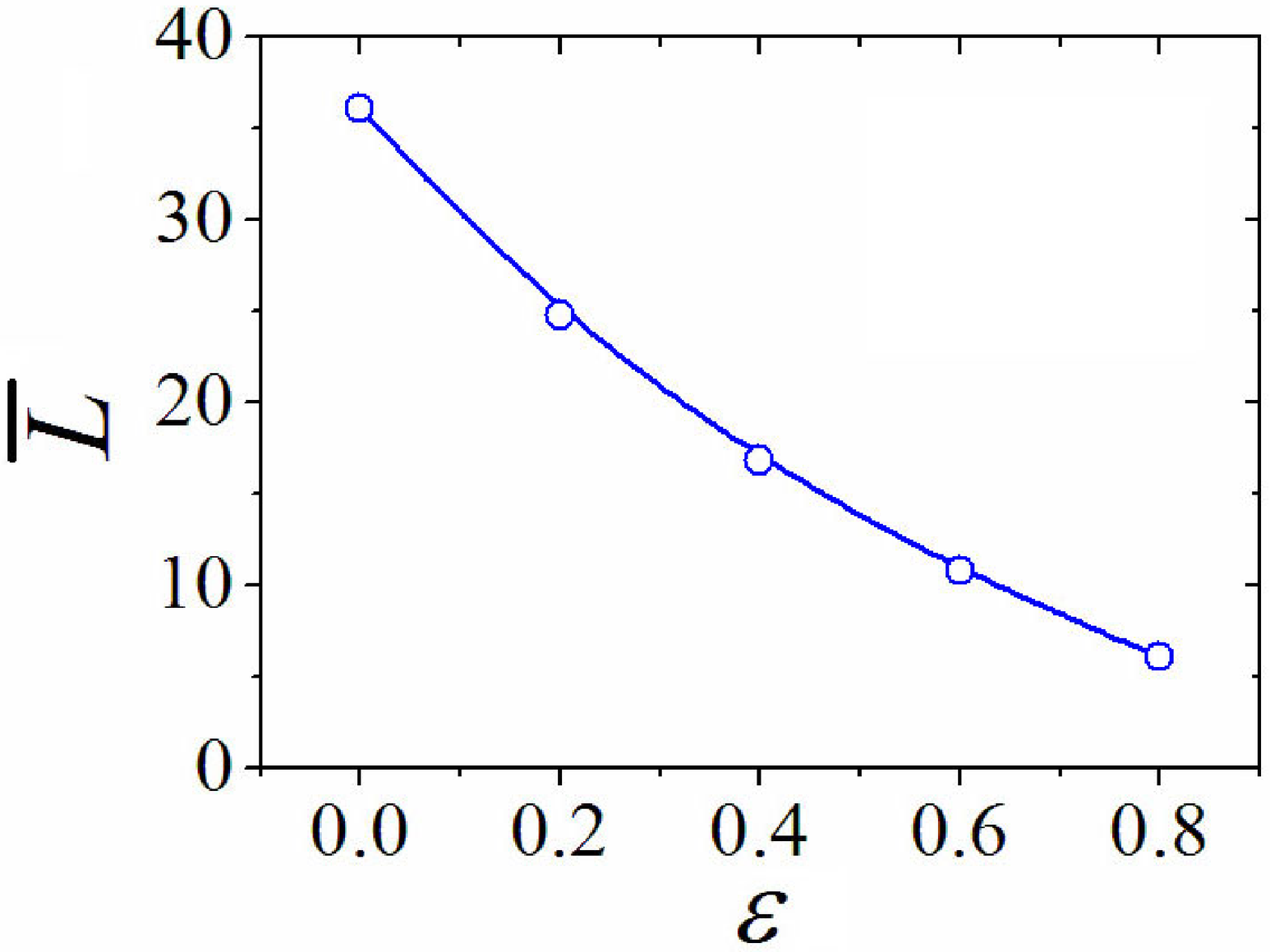}}
\caption{(Color online) For the same case of Fig.~\ref{totalwave}: (a) width $\overline{L}(t)$ for $\protect\varepsilon %
=0$~(black squares), $0.4$~(red circles) and $0.8$~(green triangles); (b)
$\overline{L}(t=200)$ as a function of the disorder strength $\protect\varepsilon $.}
\label{Output}
\end{figure}

We here consider an initially localized imbalance excitation on top of an
otherwise perfectly balanced system, i.e., $a_{n}=b_{n}=1/\sqrt{2}$ for all $%
n$, except for $a_{0}=1$ and $b_{0}=0$ at $n=0$. In the following, we focus
on $\eta =0$ and fix $P=0.1<P_{\mathrm{cr}}$~(note that, for $P>P_{\mathrm{cr%
}}$, the balanced background would be unstable). To study more accurately the effect of the
disorder on the imbalance transport, we analyze a large
number, $K=500$, of random realizations. Figure~\ref{totalwave} shows the
average spatial profile of the imbalanced perturbation, $\overline{S_{n}^{z}}%
(t)=K^{-1}\sum_{s=1}^{K}|S_{n}^{z(s)}(t)|$, where $S^{z(s)}$ is the
imbalance distribution of the $s$-th realization. When $\varepsilon =0$, the
system is homogeneous, and the initial perturbation propagates
ballistically, as seen in Fig.~\ref{totalwave}~(a). In contrast, at $%
\varepsilon \neq 0$, the expansion from the input defect at $t=0$ is no
longer ballistic, the initial imbalanced perturbation localizing around the
center, as shown in Figs.~\ref{totalwave}(b) and~(c). The respective
imbalance profile at $t=200$ is displayed in Fig.~\ref{totalwave}(d). At
sufficiently large $\varepsilon $, the imbalanced perturbation remains
exponentially localized, resembling Anderson localization. As shown in
Fig.~\eqref{Output}, localization is best quantified by monitoring the
mean size of the imbalanced perturbation, $\overline{L}(t)=K^{-1}%
\sum_{s=1}^{K}L^{(s)}(t)$, with
\begin{equation}
L^{(s)}(t)=\sqrt{\frac{\sum_{n}n^{2}|S_{n}^{z(s)}|}{\sum_{n}|S_{n}^{z(s)}|}}
\end{equation}%
being the width of the imbalance distribution of the $s$-th realization. The
localization length reduces to few wires when $\varepsilon >0.5$.


\section{Conclusions}

\label{sec:Conclusions} In summary, dipolar Bose-Einstein condensates in an
array of double-well potentials offer a simple setup which makes it possible
to employ the motional degrees of freedom for realizing an effective
mean-field transverse Ising model with peculiar inter-layer interactions.
The system gives rise to an anomalous first-order ferromagnetic-antiferromagnetic
transition, as well as to nontrivial
phases induced by frustration. As the parameters can be easily modified in real time, the
introduced setup allows as well the study of Kibble-Zurek
defect-formation. Furthermore, random occupation in each layer results in random
Ising interactions and random effective local transverse fields, which may
be employed to controllably study Anderson-like localization of imbalanced
perturbations.


\begin{acknowledgments}
This work was supported by National Natural Science Foundation of China through grants No. 11575063, No. 11374375 and No. 11574405, by the German-Israel Foundation through grant No.
I-1024-2.7/2009, and by the Tel Aviv University in the framework of the ``matching" scheme for a postdoctoral fellowship of Y.L. The work of B.A.M, is supported, in part, by the joint program in
physics between the National Science Foundation (US) and Binational Science Foundation (US-Israel), through Grant
No.2015616.
L.S. thanks the support of the DFG (RTG 1729, FOR 2247).
\end{acknowledgments}


\bibliographystyle{plain}
\bibliography{apssamp}

\begin{thebibliography}{99}
\bibitem{footnote-quantum} The realization of a coherent Josephson-like coupling between the sites demands a tight-enough axial potential
such that quantum or thermal phase fluctuations along the quasi-one-dimensional wires, and hence between the sites,
can be neglected.  Moreover, due to the assumed weakly-interacting nature of the system,
quantum fluctuations of inter-site Bogoliubov excitations are not expected to affect the qualitative nature of the phases or phase transitions discussed in this paper,
but rather lead to slight displacements of the phase boundaries.
\end{thebibliography}

\end{document}